\documentclass{aastex63}
\received{January 1, 2018}
\revised{January 7, 2018}
\accepted{\today}

\shorttitle{Rapid Flux Angle Measurements}
\shortauthors{Vech et al.}

\begin{document}


\title{Kinetic Scale Spectral Features of Cross Helicity and Residual Energy in the Inner Heliosphere}

\author{Daniel Vech}
\affiliation{Climate and Space Sciences and Engineering, University of Michigan, Ann Arbor, MI, USA}
\affiliation{Laboratory for Atmospheric and Space Physics, University of Colorado, Boulder, CO, USA\\}

\author{Justin C. Kasper}
\affiliation{Climate and Space Sciences and Engineering, University of Michigan, Ann Arbor, MI, USA}
\affiliation{Smithsonian Astrophysical Observatory, Cambridge, MA, USA}

\author{Kristopher G. Klein}
\affiliation{Lunar and Planetary Laboratory, University of Arizona, Tucson, AZ, USA\\}

\author{Jia Huang}
\affiliation{Climate and Space Sciences and Engineering, University of Michigan, Ann Arbor, MI, USA}

\author{Michael L. Stevens}
\affiliation{Smithsonian Astrophysical Observatory, Cambridge, MA, USA}

\author{Christopher H.K. Chen}
\affiliation{School of Physics and Astronomy, Queen Mary University of London, London E1 4NS, UK}

\author{Anthony W. Case}
\affiliation{Smithsonian Astrophysical Observatory, Cambridge, MA, USA}

\author{Kelly Korreck}
\affiliation{Smithsonian Astrophysical Observatory, Cambridge, MA, USA}

\author{Stuart D. Bale}
\affiliation{Space Science Laboratory, University of California Berkeley, , Berkeley, CA, USA\\}

\author{Trevor A. Bowen}
\affiliation{Space Science Laboratory, University of California Berkeley, , Berkeley, CA, USA\\}

\author{Phyllis L. Whittlesey}
\affiliation{Space Science Laboratory, University of California Berkeley, , Berkeley, CA, USA\\}

\author{Roberto Livi}
\affiliation{Space Science Laboratory, University of California Berkeley, , Berkeley, CA, USA\\}

\author{Davin E. Larson}
\affiliation{Space Science Laboratory, University of California Berkeley, , Berkeley, CA, USA\\}

\author{David Malaspina}
\affiliation{Laboratory for Atmospheric and Space Physics, University of Colorado, Boulder, CO, USA\\}

\author{Marc Pulupa}
\affiliation{Space Science Laboratory, University of California Berkeley, , Berkeley, CA, USA\\}

\author{John Bonnell}
\affiliation{Space Science Laboratory, University of California Berkeley, , Berkeley, CA, USA\\}

\author{Peter Harvey}
\affiliation{Space Science Laboratory, University of California Berkeley, , Berkeley, CA, USA\\}

\author{Keith Goetz}
\affiliation{School ofPhysics and Astronomy, University of Minnesota, Minneapolis, Minnesota, USA\\}

\author{Thierry Dudok de Wit}
\affiliation{LPC2E, CNRS and University of Orl\'{e}ans, Orl\'{e}ans, France\\}

\author{Robert MacDowall}
\affiliation{NASA Goddard Space Flight Center, Greenbelt, MD, USA\\}

\date{\today}

\begin{abstract}

In this Paper, we present the first results from the Flux Angle operation mode of the Faraday Cup instrument onboard Parker Solar Probe. The Flux Angle mode allows rapid measurements of phase space density fluctuations close to the peak of the proton velocity distribution function with a cadence of 293 Hz. This approach provides an invaluable tool for understanding kinetic scale turbulence in the solar wind and solar corona. We describe a technique to convert the phase space density fluctuations into vector velocity components and compute several turbulence parameters such as spectral index, residual energy and cross helicity during two intervals the Flux Angle mode was used in Parker Solar Probe's first encounter at 0.174 AU distance from the Sun.\\

\end{abstract}



\section{Introduction} \label{sec:intro}
The solar wind is a hot, tenuous plasma propagating away from the Sun's surface, which is ubiquitously observed in a turbulent state \citep{coleman1968turbulence}. Turbulence in the solar wind is modelled as a cascade of energy from the outer scales to the much smaller dissipative scales through an inertial range. In the inertial range the velocity and magnetic fluctuations are largely perpendicular to the local magnetic field direction and the spectral index of the power spectra of the magnetic and velocity fluctuations are close to -5/3 and -3/2, respectively \citep{coleman1968turbulence,matthaeus1982measurement, podesta2007spectral, boldyrev2011spectral}. Below this range roughly coincident with the convected ion kinetic scales the magnetic energy spectrum steepens and Alfv\'enic turbulence undergoes a transition into dispersive kinetic Alfv\'en waves (KAW) \citep{bale2005measurement, chen2013nature}. Between ion and electron scales the spectral index of the magnetic fluctuations is typically between -2 and -4 \citep{leamon1998observational, smith2006dependence,  alexandrova2009universality, matteini2016electric}.

In contrast to magnetic fields, the power spectrum of velocity fluctuations in the kinetic range is much less understood due largely to the fact that high cadence plasma moment measurements in the solar wind became only recently available. Studies based on Spektr-R data (proton moments with 31 ms cadence measured by six Faraday Cups) presented the first results on the high frequency part (up to 2 Hz) of the velocity power spectrum including break frequency and spectral indices \citep{vsafrankova2013fast, vsafrankova2013ion, vsafrankova2016power, riazantseva2017variety}. Unfortunately the lack of an operating magnetic field instrument of Spektr-R made it impossible to study correlation between high frequency velocity and magnetic fluctuations and compute cross helicities and residual energies.

Cross helicity is defined as $\sigma_c = ({E}^+ - E^-)/(E^+ + E^-)$ where $E^{\pm}$ corresponds to the power spectra of the Elsasser variables $\bf{z}^\pm = \delta \bf{v} \pm \delta \bf{b}/ \sqrt{\mu_0 \rho}$ where $\delta \bf{v}$ and $\delta \bf{b}$ are the fluctuations of the velocity and magnetic fields in Alfv\'{e}n units, respectively and $\rho$ is the mean mass density of protons \citep{wicks2013correlations, chen2013residual}. Cross helicity is normalized in such a way that it is 1 and -1 for anti-Sunward and Sunward propagating waves, respectively. Cross helicity is conserved in the absence of dissipation and corresponds to the linkages between lines of vorticity and magnetic field lines, both of which are frozen to the fluid flow in the absence of dissipation \citep{chandran2008strong}. In addition to the numerous statistical studies at 1 AU \citep[e.g.][]{wicks2013correlations, chen2013residual}, the radial dependence of $\sigma_c$ was investigated on MHD scales with Helios data \citep{roberts1987origin, bruno1991origin, grappin1990origin, bruno1991origin, bruno1993cross, bruno1997inward}. These studies found that $\sigma_c$ decreases as the solar wind propagates away from the Sun, however the physical mechanisms responsible for this feature are debated. For example, \cite{bruno1991origin} and \cite{bruno1993cross} suggested that the decrease of $\sigma_c$ is driven by the interaction of Alfv\'{e}nic fluctuations with static structures or magnetosonic perturbations, which results in a decrease of $\bf{z}^+$ component rather than an increase  of $\bf{z}^-$. Several other studies suggested that shear and expansion causes the decrease of $\sigma_c$ with increasing radial distance \citep[e.g.][]{roberts1987origin, oughton1995linear}.

Residual energy is the difference between the kinetic and magnetic energy  $\sigma_r = (E^v - E^b)/(E^v + E^b)$. Unlike for pure Alfv\'{e}n waves \citep{alfven1942existence} where the energy of velocity and magnetic fluctuations are in equipartition, in the solar wind the magnetic energy is typically larger than the energy of velocity fluctuations \citep{wicks2013correlations, chen2013residual}. The origin of this difference is a matter of considerable debate: potential explanations include the role of magnetic structures with solar origin and local generation of residual energy by counterpropagating Alfv\'{e}n wave packets \citep{wang2011residual, boldyrev2012residual, bowen2018impact}.



Understanding the scaling of $\sigma_r$ and $\sigma_c$ in the kinetic-scale solar wind fluctuations is fundamentally important for describing heating and dissipation in the solar wind, solar corona and plasma systems more generally. Previous turbulence models \citep[e.g.][]{boldyrev2006,matthaeus2008rapid} described the coupling between velocity and magnetic field fluctuations in the inertial range, however the main assumptions underlying these models are violated at the kinetic scales where the MHD approximation breaks down and the quadratic integral invariants are no longer retained \citep{matthaeus1982measurement}. To the best of our knowledge no theory exists that describes the correlation between velocity and magnetic fluctuations in the kinetic range. The first attempt to measure $\sigma_c$ and $\sigma_r$ in the kinetic range was presented by \cite{parashar2018kinetic} using Magnetospheric Multiscale Mission (MMS) data. They found that $\sigma_r$ and $\sigma_c$ converge to 1 and 0, respectively from the inertial range to the smallest observable scales (20-40 km). The loss of alignment between $\delta \bf{v}$ and $\delta \bf{b}$ (quantified by the metric $cos(\theta) =\sigma_c / \sqrt{(1-\sigma_r^2)} \approx$ 0) was explained by the demagnetization of protons. 

The Faraday Cup (SPC) instrument \citep{kasper2016solar, Case:SPC} onboard NASA's Parker Solar Probe (PSP) \citep{fox2016solar} is equipped with a novel Flux Angle (FA) operation mode that allows rapid measurements of the phase space density fluctuations with an unprecedented 293 Hz cadence providing a new tool to understand kinetic scale turbulence in the solar wind and solar corona. SPC was operated in FA mode twice for approximately a total of 110 seconds during the first perihelion of PSP on 4th November 2018 and captured the fine structure of a magnetic switchback. Magnetic switchbacks are one of the most prominent features of the solar wind in the inner heliosphere; they are characterized by short, large amplitude velocity enhancements that are accompanied with 90-180$^{\circ}$ rotation of the magnetic field \citep{gosling2011pulsed, horbury2018short, Horbury:prep}. These structures might be direct signatures of impulsive chromospheric or coronal energy release \citep{horbury2018short, Bale:prep}.

In this Paper, we present the first results from the FA operation mode of SPC and study $\sigma_c$ and $\sigma_r$ in the kinetic range of the turbulent cascade. Our study complements the ones by \cite{Chen:prep} and \cite{Parashar:prep}, which focus on magnetic and velocity fluctuations on MHD scales in the inner heliosphere. In Section 2 we describe the conversion of phase space fluctuations into vector velocity fluctuations, with particular emphasis on the underlying assumptions and limitations of the data product. In Section 3, we discuss the properties of kinetic scale turbulence in the observed magnetic switchback such as spectral index of the power spectrum, residual energy and cross helicity. Finally, Section 4 contains a summary and a discussion of the results.


\section{Method} \label{sec:method}


The Faraday Cup instrument of PSP measures fluxes and flow angles as a function of energy from 50 eV/q to 8 keV/q for ions \citep{kasper2016solar, Case:SPC} based on the currents detected by four collector plates. In typical operation mode SPC scans through 128 energy per charge windows in 0.87 seconds (1.14 Hz); higher cadence data products ($\sim$5-19.6 Hz) are available for shorter intervals as well. In FA mode SPC measures a single energy/charge window near the center of the proton velocity distribution function (VDF) with 293 Hz cadence.

Figure 1 shows an overview of the components of the magnetic field (293 Hz cadence based on fluxgate magnetometer data, \cite{bale2016fields}) and velocity for a 3 hr period starting on November 4th 2018 14:00:00 UT when PSP was approximately at 0.174 AU distance from the Sun. The vector components are in the RTN coordinate system where R points radially outward from the Sun, N is along the ecliptic North and T completes the right hand coordinate system. A magnetic switchback was observed from 15:24:01 to 15:43:07 and was accompanied with a sudden reversal of the radial magnetic field component and enhanced (toward the positive T direction) tangential velocity component. The duration of the studied magnetic switchback (e.g. interval with B$_R >$ 0 nT) is approximately 19 minutes, which is considered to be an above average structure \citep[see][]{Kasper2019:prep}. In Figure 1 the shaded region marks the two intervals where SPC was operated in FA mode between 15:31:54-15:32:53 and 15:33:30-15:34:22 UT.

Figure 2 shows 15-second averages of proton VDFs before each FA mode interval where x-axis is the phase speed and y-axis is the phase space density ($P$) in arbitrary units (for the conversion of the axes see \cite{Case:SPC}). The FA mode achieves unprecedented temporal resolution by scanning a single window in phase space near the peak of the VDF, which are marked with blue (446-457 km/s) and red (426-437 km/s) for the first and second FA mode intervals, respectively. Significant changes in the solar wind parameters shift the VDF hence the blue and red regions do not align with the peak, which makes the interpretation of the FA mode measurements more complicated. To ensure that the FA mode interval is not affected by those large changes, we studied the variability of the solar wind parameters and compared 15-second averages of the solar wind speed (V$_{sw}$), core proton density ($n_p$), core thermal velocity (V$_{th}$), ratio of thermal to magnetic pressure ($\beta_p$) and Alfv\'{e}n speed (V$_A$), which are summarized in Table 1. The solar wind parameters were very steady during the studied periods and none of them show variations of more than 5\% suggesting that SPC measured approximately the same part of the VDF throughout in the FA mode intervals. We note that the proton core temperature anisotropy ($T_{\perp} / T_{||}$ estimated with 10-sec cadence, for details see \cite{Huang:prep}) was in the range of 0.96-1.07 for both intervals.

\begin{table}[h!]
\centering
    \label{tab:table1}
    \begin{tabular}{l|r|r|r|r} 
      Parameter & Before \#1 & After \#1 & Before \#2 & After \#2 \\
      \hline
      $V_{sw}$ [km/s] & 423.5 $\pm$ 5.3 & 415.9 $\pm$ 12.1 & 406.7 $\pm$ 4.1 & 421 $\pm$ 4.2\\
      $n_{p}$ [cm$^{-3}$] & 231.5 $\pm$ 16.9 & 230.1 $\pm$ 19.4 & 241.3 $\pm$ 9.8 & 231.7 $\pm$ 16.7\\
      $V_{th}$ [km/s] & 78.0 $\pm$ 5.7 & 82.7 $\pm$ 5.9 & 85.9 $\pm$ 3.4 & 81.7 $\pm$ 5.7\\
      $\beta_p$ & 0.6 $\pm$ 0.17 & 0.63 $\pm$ 0.16 & 0.71 $\pm$ 0.08 & 0.77 $\pm$ 0.2\\
      $V_A$ [km/s] & 102.1 $\pm$ 5.1 & 104.7 $\pm$ 4.6 & 102.3 $\pm$ 2.2 & 96.9 $\pm$ 3.5\\
    \end{tabular}
        \caption{15-second averages and standard deviations of solar wind parameters before and after each FA mode interval.}
\end{table}


The goal of the subsequent analysis is to define a fitting procedure converting the FA mode data into vector velocity components and to estimate the potential noise contribution from $n_p$ and $V_{th}$ fluctuations. Our approach is the following: full proton VDFs from 15:30:54 to 15:35:22 UT (starting 60 seconds before the first and ending 60 seconds after the second FA mode interval) were selected where plasma moments (including $V_{sw}$, $n_p$ and $V_{th}$) were available. Three linear fits were used to estimate the scaling of $P$ with $V_{sw}$, $n_p$ and $V_{th}$. For the fitting, the phase space densities in the 445-457 km/s and 426-437 km/s windows were normalized by their mean values based on the entire interval ($\tilde{P}$ = $P$/$<P>$). The slopes (L$_{1,2}$) and intercepts (M$_{1,2}$) of the fits are summarized in Table 2. We note that in the studied interval $P$ varies over a relatively small range, and thus higher order fits lead to no meaningful improvement in the goodness of these fits.

\begin{table}[h!]
\centering
    \label{tab:table1}
    \begin{tabular}{c|r|r|r|c} 
      Response variable & L$_1$ & L$_2$ & M$_1$ & M$_2$ \\
      \hline
      $V_{sw}$ & 111.19 & 87.677 & 320.16 & 337.67  \\
      $V_{th}$  & -43.973 & -48.672 & 126.76 & 131.46  \\
      $n_p$  & -76.941 & -46.356  & 311.69 & 281.1 \\
    \end{tabular}
        \caption{ Fitting parameters for the 445-457 km/s (L$_1$, M$_1$) and 426-437 km/s (L$_2$, M$_2$) phase space density fluctuations, respectively.}
\end{table}


The vector velocity components in the RTN frame were obtained as

\begin{equation}
V_R = [\emph{cos}(\phi) \cdot \emph{cos}(\theta) \cdot ((L_{V_{sw 1,2}} \cdot \tilde{P}_{{1,2}}) + M_{V_{sw 1,2}} )] - V_{R_{S/C}}
\label{eqn:lambda}
\end{equation}

\begin{equation}
V_T = [\emph{cos}(\phi) \cdot \emph{sin}(\theta) \cdot ((L_{V_{sw 1,2}} \cdot \tilde{P}_{{1,2}}) + M_{V_{sw 1,2}} )] - V_{T_{S/C}}
\label{eqn:lambda}
\end{equation}

\begin{equation}
V_N = [\emph{sin}(\phi) \cdot ((L_{V_{sw 1,2}} \cdot \tilde{P}_{{1,2}}) + M_{V_{sw 1,2}} )] - V_{N_{ S/C}}
\label{eqn:lambda}
\end{equation}

\begin{equation}
\phi = \frac{2\pi}{180} \lambda [ \frac{(C + D) - (A+B)} {A + B + C + D} ]
\label{eqn:lambda}
\end{equation}

\begin{equation}
\theta = \frac{2\pi}{180} \lambda [ \frac{(A + D) - (B+C)} {A + B + C + D} ].
\label{eqn:lambda}
\end{equation}

Equation 4 and 5 are linear approximations of the flow angles where the constant $\lambda = (\pi/2) * (r_{LA}/d_{LA}) \approx 1.009$ is determined by the radius of the limiting aperture, $r_{LA}=10.86$  mm, and its axial distance from the collector plates, $d_{LA}=16.9$ mm. The denominator of Equations 4 and 5 corresponds to the sum of currents measured by the four collector plates (A, B, C, D) of SPC. As viewed from the Sun during encounter, A and B collector plates are on the "ecliptic south" while C and D are on the "ecliptic north". Equation 4 thus measures the elevation angle of the flow from the R-T plane toward the N+ direction. Similarly, the numerator of Equation 5 corresponds to currents in the "east-west" direction and thus $\theta$ is the azimuth angle of the flow, which is measured in the R-T plane from the R+ direction toward T+ direction (for details of the operation of SPC see \cite{Case:SPC}). In Equation 1-3 $\tilde{P}_{{1,2}}$ is the normalized phase space density in the 445-457 km/s and 426-437 km/s windows, V$_{i_{SC}}$ is the i$^{th}$ component of the spacecraft velocity in the RTN frame (V$_{R_{SC}}$ = 18.17 km/s, V$_{T_{SC}}$ = -90.7 km/s, V$_{N_{SC}}$ = -4.1 km/s during both intervals).

Figure 3 compares the directly measured (based on 19.6 Hz moments) and estimated (using Equation 1-5) RTN velocity components for the 445-457 km/s (a-c) and 426-437 km/s (d-f) phase space density fluctuations, respectively. The corresponding R-squared values and $1\sigma$ errors are shown in each panel; the red line is x=y. Figure 4 has the same format as Figure 3 and shows the measured and estimated $n_p$ and $V_{th}$ values. For the 445-457 km/s phase speed (Figure 4a-c) the predicted RTN velocity components are in excellent agreement with the high cadence moments. Figure 4a-b shows that the predictive power of $P$ is somewhat lower for $V_{th}$ and $n_p$ resulting in lower R-squared values. For the lower phase speed range (Figure 4d-f) we found some scattering in the $V_R$ component while the T and N velocity components are in good agreement with the measured values. In Figure 4c-d $V_{th}$ and $n_p$ are predicted with larger errors than in Figure 4a-b and the fits have the lowest R-squared.

We use our fitting technique to estimate the vector velocity fluctuations in the FA mode and compare their spectral properties to the velocity moments derived based on full VDFs. We selected 190 seconds of data between 15:28:44-15:31:54 and 15:34:33-15:37:32 (e.g. measurements right before and after the first and second FA mode intervals, respectively) when SPC measured full VDFs with 19.6 Hz cadence. The length of this interval was chosen such that it is long enough to cover the inertial range of the fluctuations but also all data points are within the magnetic switchback. The trace power spectrum of the velocity fluctuations was computed for the 19.6 Hz data in the magnetic switchback and compared to the spectrum of fluctuations derived from the FA mode data. The results in Figure 5 suggest that the two data sets have remarkably good agreement for low frequencies (below 1 Hz) for both the first (a) and second (b) FA mode intervals. We note that in the case of the 19.6 Hz data switching between neighbouring energy/charge windows during the scans may introduce some noise, which results in a higher noise floor as compared to the FA mode data. 

Figure 4 indicates that the $P$ fluctuations are correlated with $V_{th}$ and $n_p$ as well, which may introduce noise in the velocity power spectra in Figure 5. To quantify this effect, we used the first FA mode interval and calculated the RTN velocity components, $V_{th}$ and $n_p$. Each parameter was normalized to its median value (e.g. each velocity component separately) and then the power spectra were computed. In Figure 6 it can be seen that the wave power of the trace velocity fluctuations is three orders of magnitude larger than the wave power of $n_p$ and $V_{th}$. This significant difference is explained by the fact that the flow angle shows rapid and large amplitude fluctuations (as expected for a highly Alfv\'{e}nic flow) while $n_p$ and $V_{th}$ change much more slowly. Based on these results we suggest that the velocity spectra has negligible noise contribution from changes in $n_p$ and $V_{th}$.




\begin{figure}
    \centering\includegraphics[width=0.7\linewidth]{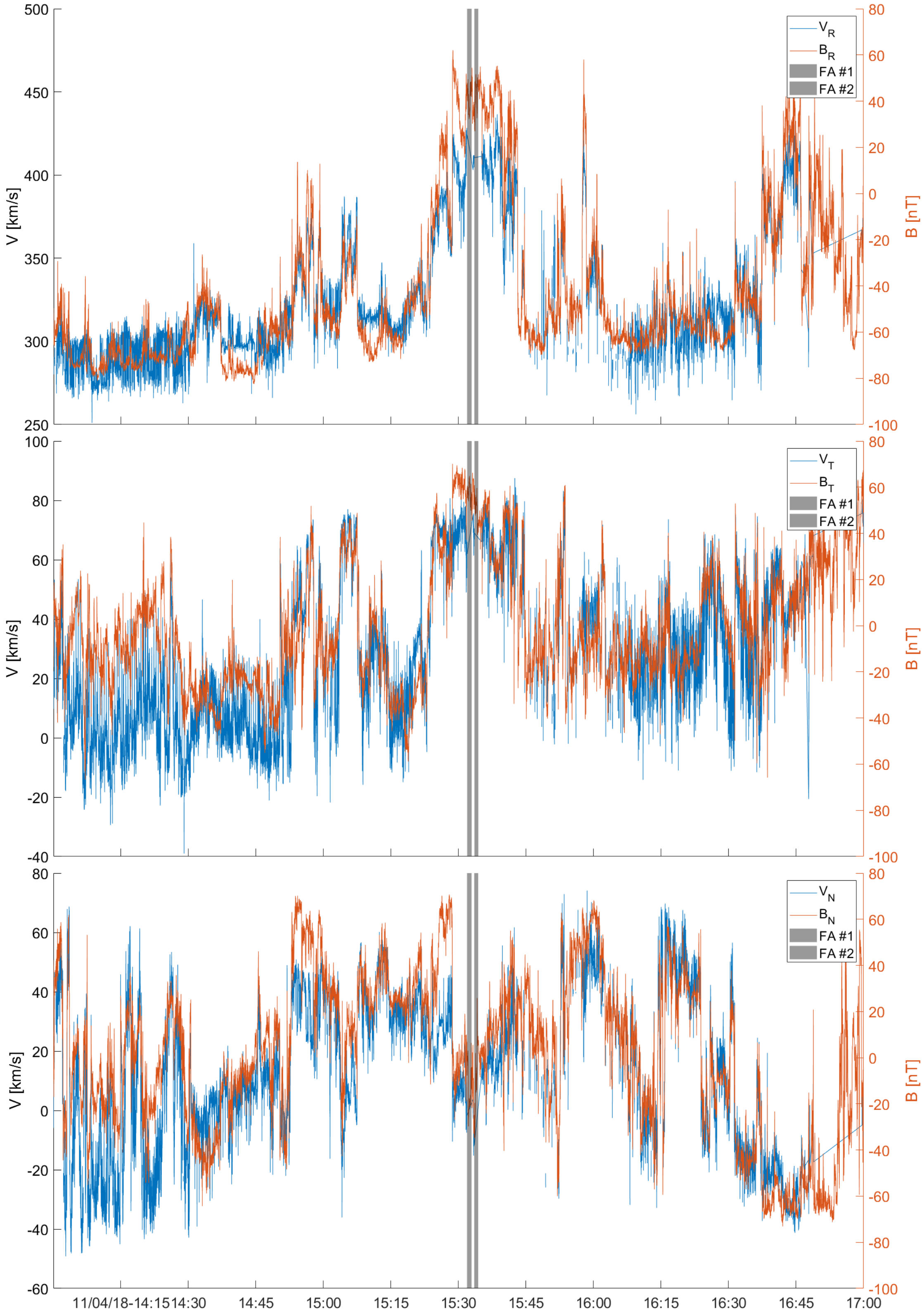}
\caption{Overview of the magnetic and velocity components in a 3hr interval centered at the FA mode data.}
  \label{fig:1}
\end{figure}

\begin{figure}
    \centering\includegraphics[width=0.7\linewidth]{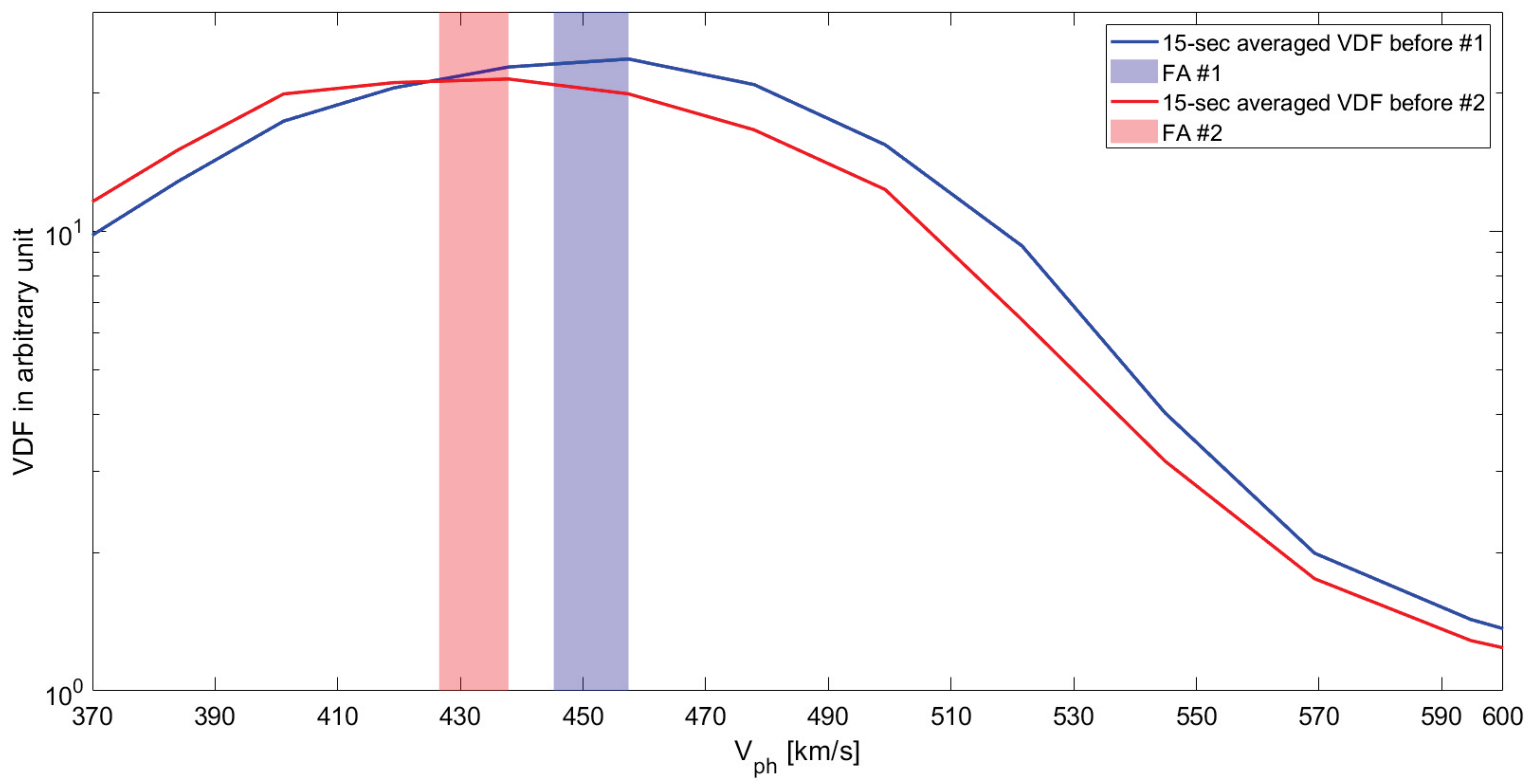}
\caption{15-second averages of full VDFs before each FA mode interval, respectively. The shaded areas mark the range in phase speed, which are measured in the FA mode.}
  \label{fig:2}
\end{figure}

\begin{figure}
    \centering\includegraphics[width=1\linewidth]{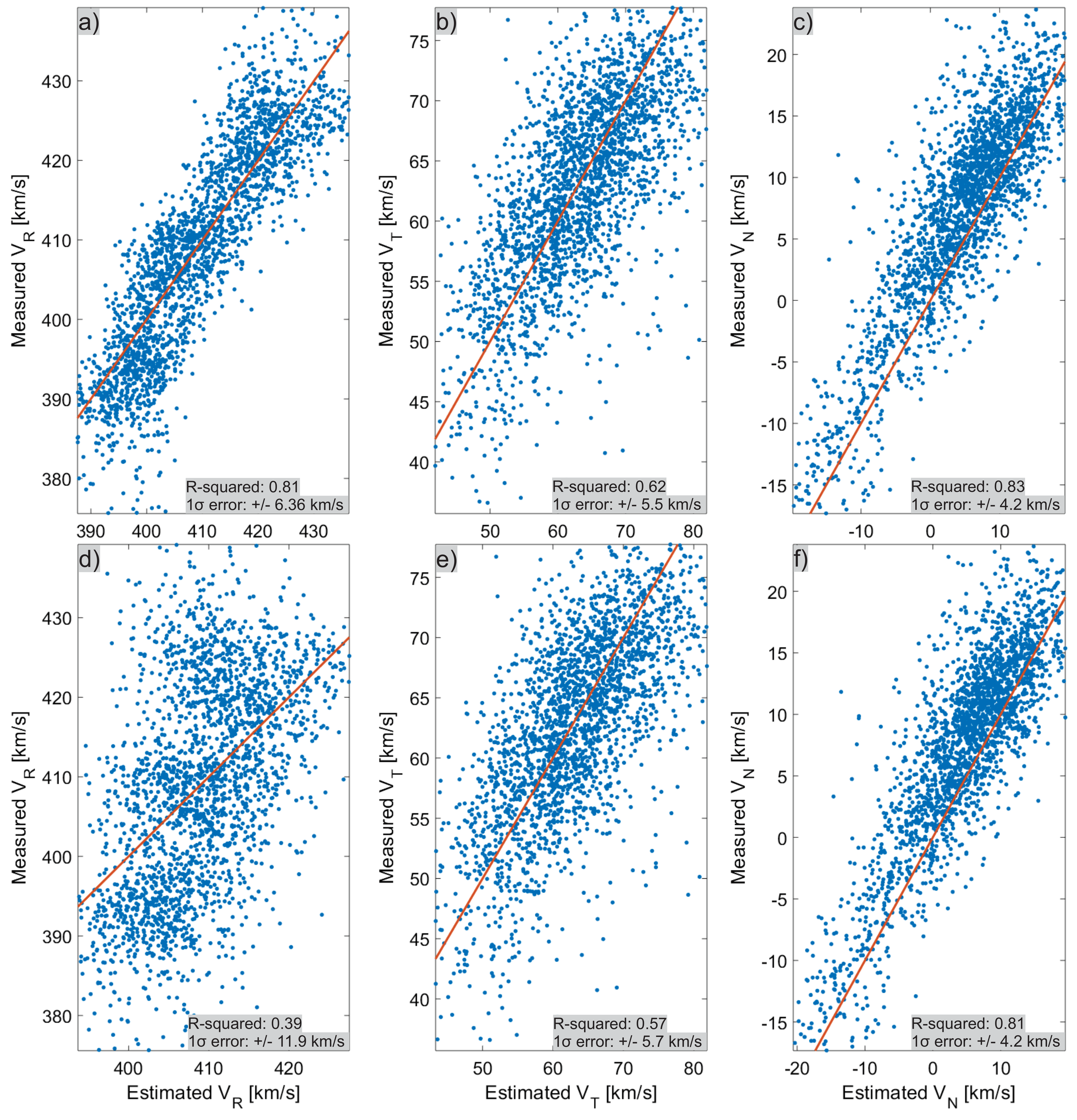}
\caption{Comparison of the measured (based on 19.6 Hz moments) and estimated (Equation 1-5) RTN velocity components for the 445-457 km/s (a-c) and 426-437 km/s (d-f) phase space density fluctuations, respectively.}
  \label{fig:3}
\end{figure}

\begin{figure}
    \centering\includegraphics[width=1\linewidth]{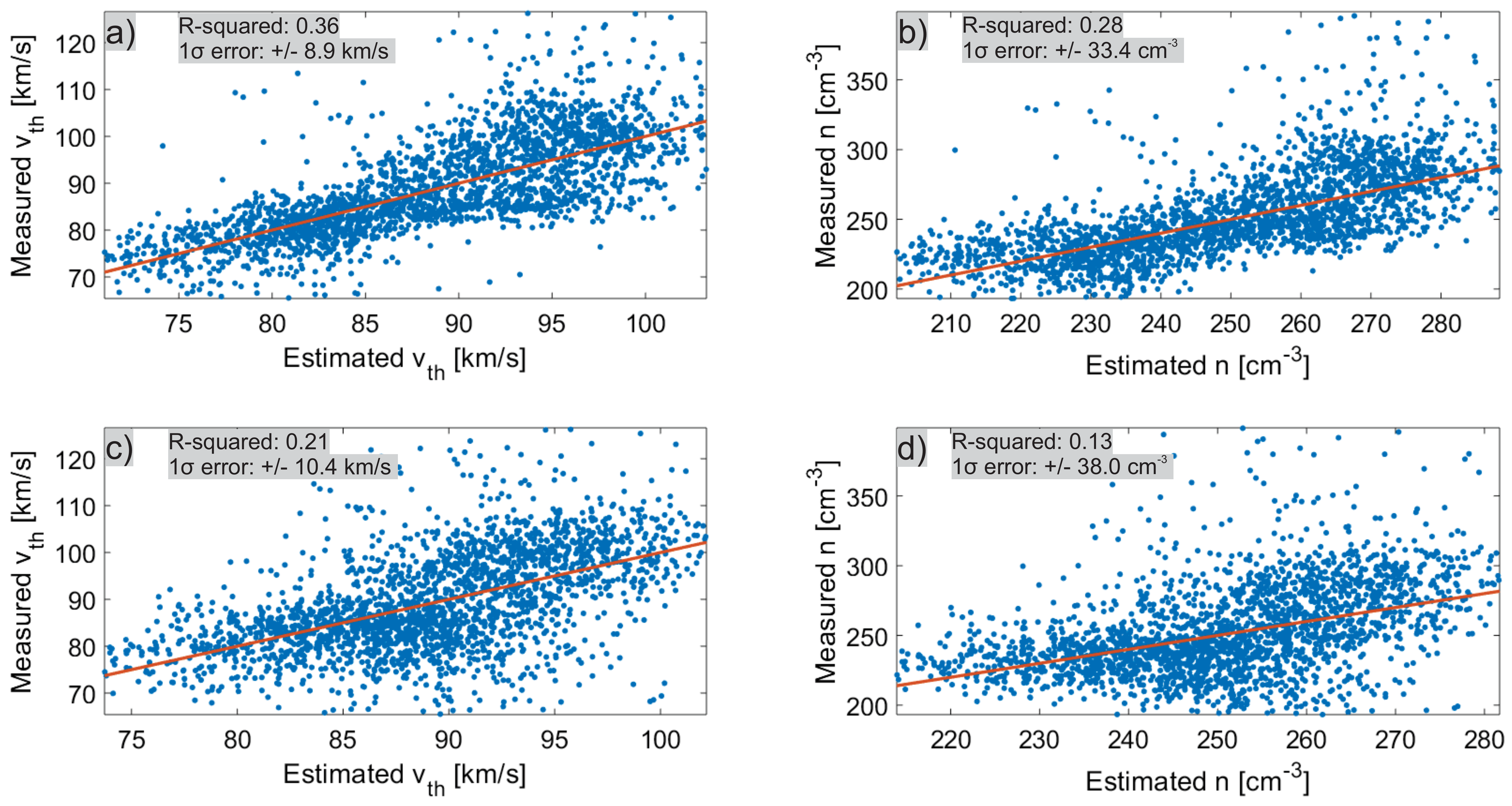}
\caption{Comparison of the measured (based on 19.6 Hz moments) and estimated $n_p$ and $V_{th}$ for the 445-457 km/s (a-b) and 426-437 km/s (c-d) phase space density fluctuations, respectively.}
  \label{fig:4}
\end{figure}

\begin{figure}
    \centering\includegraphics[width=0.7\linewidth]{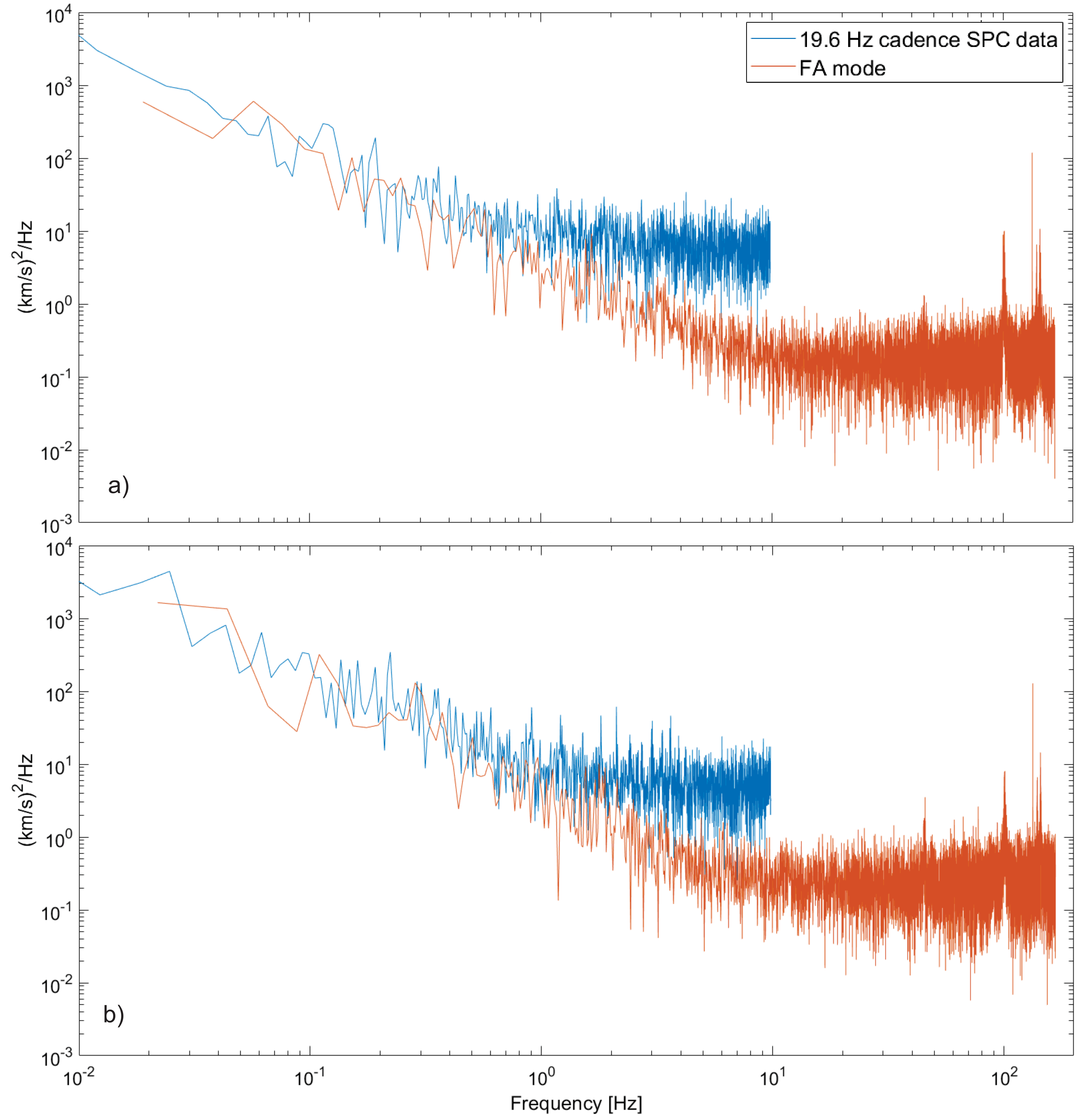}
\caption{Comparison of the trace power spectrum of velocity fluctuations for the 19.6 Hz cadence data when full VDFs were measured and the FA mode data in interval \#1 and \#2. For frequencies below 1 Hz the FA mode data shows remarkably good agreement with the 19.6 Hz data for both intervals.}
  \label{fig:5}
\end{figure}

\begin{figure}
    \centering\includegraphics[width=0.7\linewidth]{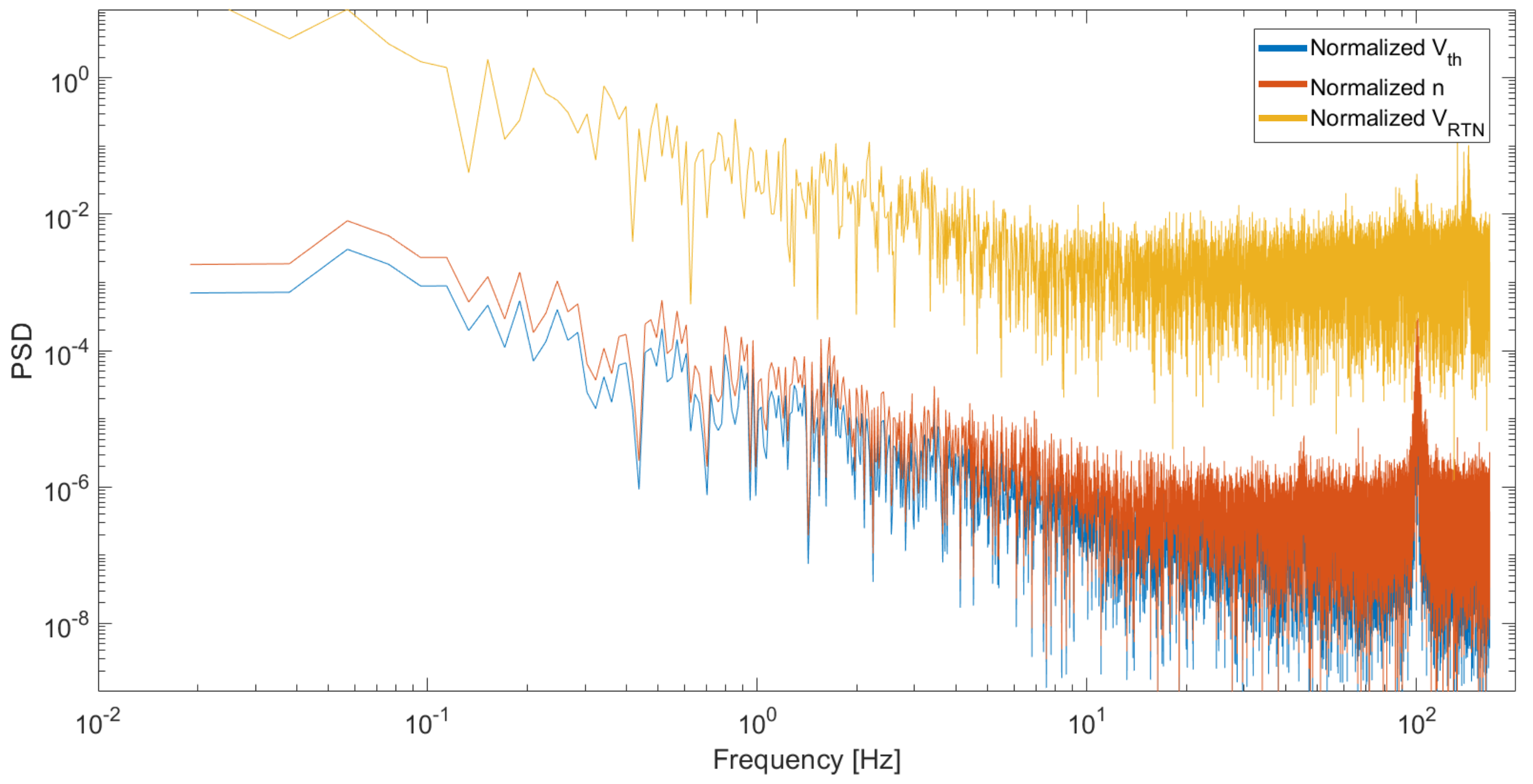}
\caption{Comparison of the power spectrum of the normalized $V_{th}$, $n_p$ and trace velocity fluctuations during the first FA mode interval.} 
  \label{fig:6}
\end{figure}


\section{Spectral features of kinetic scale turbulence} \label{sec:spectral}

Figure 7a and b show the power spectrum of the trace velocity and magnetic fluctuations. The magnetic field fluctuations were converted into Alfv\'{e}n units normalizing by ($\mu_0 \rho$)$^{1/2}$. The vertical lines mark the scale of the convected ion inertial length ($V_{sw}/2\pi d_i$) and proton gyoradius ($V_{sw}/2\pi \rho_i$) where $d_i = c/\omega_p$ and $\rho_i = m v_{\perp}/qB $. The black and green dots show the V and B-field spectral indices based on a fitting window, which has a size of a factor of 3.7; the dots are placed at the center of each fitting window.

In the inertial range (0.1-1 Hz) the spectral indices of the velocity and magnetic fluctuations are -1.51 / -1.61 and -1.60 / -1.74 for the first and second intervals, respectively. These values are close to the observations at 1 AU where magnetic field spectrum is typically steeper than the velocity \citep{boldyrev2011spectral, chen2013residual, bowen2018impact}. The ion-scale spectral break of the magnetic field power spectrum is approximately 5 and 2 Hz in the first and second intervals, respectively, which are at least a factor of six larger than the typical values at 1 AU ($\approx$ 0.3 Hz, \cite[e.g.][]{markovskii2008statistical,vech2017nature}). This suggests again that the FA mode is essential to study the $\delta \textbf{v}$-$\delta \textbf{b}$ coupling in the kinetic range since velocity fluctuations at these scales are not measured with other operation modes of SPC.

The V and B-field spectral indices show good correlation in the inertial range; at kinetic scales the B-field spectral index is around -2.5 and -3, which is similar to the observations at 1 AU \citep[e.g.][]{alexandrova2009universality,leamon1998observational}, in contrast at kinetic scales we find no signatures of spectral steepening in the V-field power spectrum. Previous studies found that the power spectrum of ion fluxes show very wide range of features: \cite{riazantseva2017variety} categorized power spectrums of ion fluxes into five groups using Spektr-R data at 1 AU. The most frequently occurring spectra (50\% of the cases) showed two slopes and one break point between them at ion-scale, the second most frequent class (32\%) showed flattening in the vicinity of the break. In contrast, 6.3\% of power spectras did not show steepening at kinetic scales at all. \cite{riazantseva2017variety} did not find clear trend (such as $V_{sw}$ or $\beta_p$ dependence) in the underlying solar wind parameters that may explain this feature. Based on previous studies \citep[e.g.][]{chen2017nature} we expect the steepening of the velocity spectra and it is possible that the noise floor of the FA mode data is not low enough to measure such a break scale.

\begin{figure}
    \centering\includegraphics[width=0.7\linewidth]{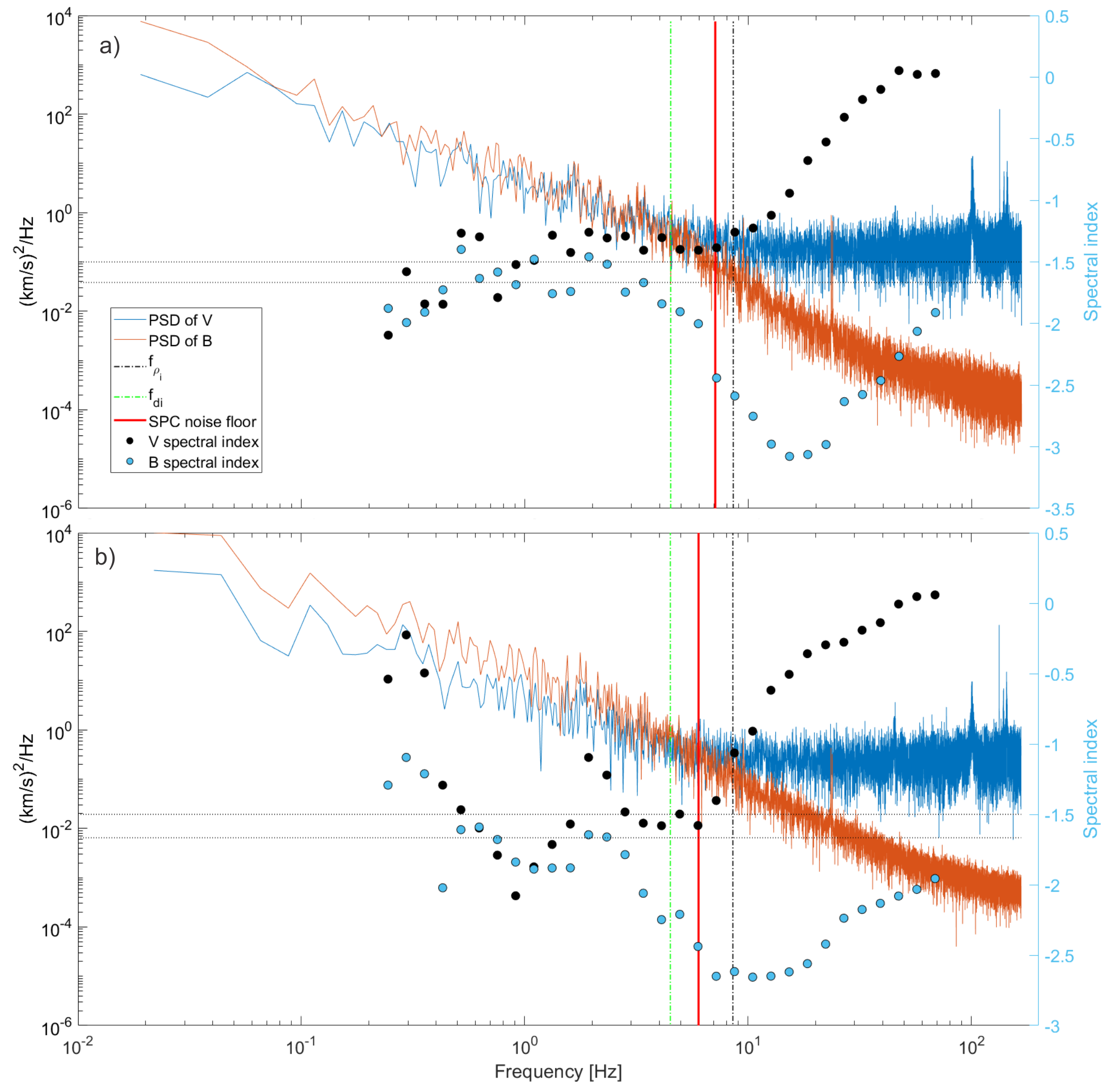}
\caption{Power spectrums of velocity and magnetic field fluctuations during the first (a) and second (b) FA mode intervals, respectively.}
  \label{fig:2}
\end{figure}

Figure 8 shows the normalized cross helicity, residual energy and cosine of the alignment angle. In Figure 8a, the normalized cross helicity shows some fluctuations in the inertial range ($\sigma_c \approx$ 0.4), which is followed by a sudden decrease near the ion-scale spectral break and convergence to 0 at kinetic scales. In Figure 8b, the magnetic energy is larger than the energy of velocity fluctuations in the inertial range and $\sigma_r$ increases gradually toward kinetic scales. Finally, Figure 8c suggests that the magnetic and velocity fluctuations are aligned in the inertial range and $cos(\theta) = \sigma_c/\sqrt{(1-\sigma_r^2)} = 0.5$, however this alignment drops at approximately 1.4 Hz ($\approx$ 3.1d$_i$), which is comparable to the values found by \cite{parashar2018kinetic} in the solar wind (4.4d$_i$) at 1 AU and in the terrestrial magnetosheath (6.5d$_i$). Disruption of current sheets with the size of a few d$_i$ may affect the turbulent cascade and lead to the lack of alignment between $\delta$\textbf{v} and $\delta$\textbf{b} \cite[see][]{mallet2017disruption, Loureiro_2017, vech2018magnetic}. Another explanation for the loss of alignment is that the turbulence transitions into kinetic Alfv\'{e}n range where the polarisation of the fluctuations changes and the alignment between $\delta \textbf{b}$ and $\delta \textbf{v}$ cease to exist \citep[e.g.][]{schekochihin2009astrophysical}.

The sudden decrease of the cosine of the alignment angle in Figure 8c is close to the flattening of the proton velocity spectra hence we used the following test to quantify the effect of noise. An artificial test velocity data ($\textbf{V}_{test}$) was computed by adding Gaussian noise to the magnetic field measurements. The amplitude of the noise was empirically chosen such that the trace power spectra of $\textbf{V}_{test}$ is in good agreement with the real one in Figure 7. We calculated $cos(\theta)$ using $\textbf{V}_{test}$ and compared it to the real measurements. We found that in the artificial test data the alignment drops to zero at a factor of 3 times higher frequency than the real measurements therefore we suggest that the observed changes of $cos(\theta)$ near the break scale are primarily physical and not caused by Gaussian noise.

\begin{figure}
    \centering\includegraphics[width=0.7\linewidth]{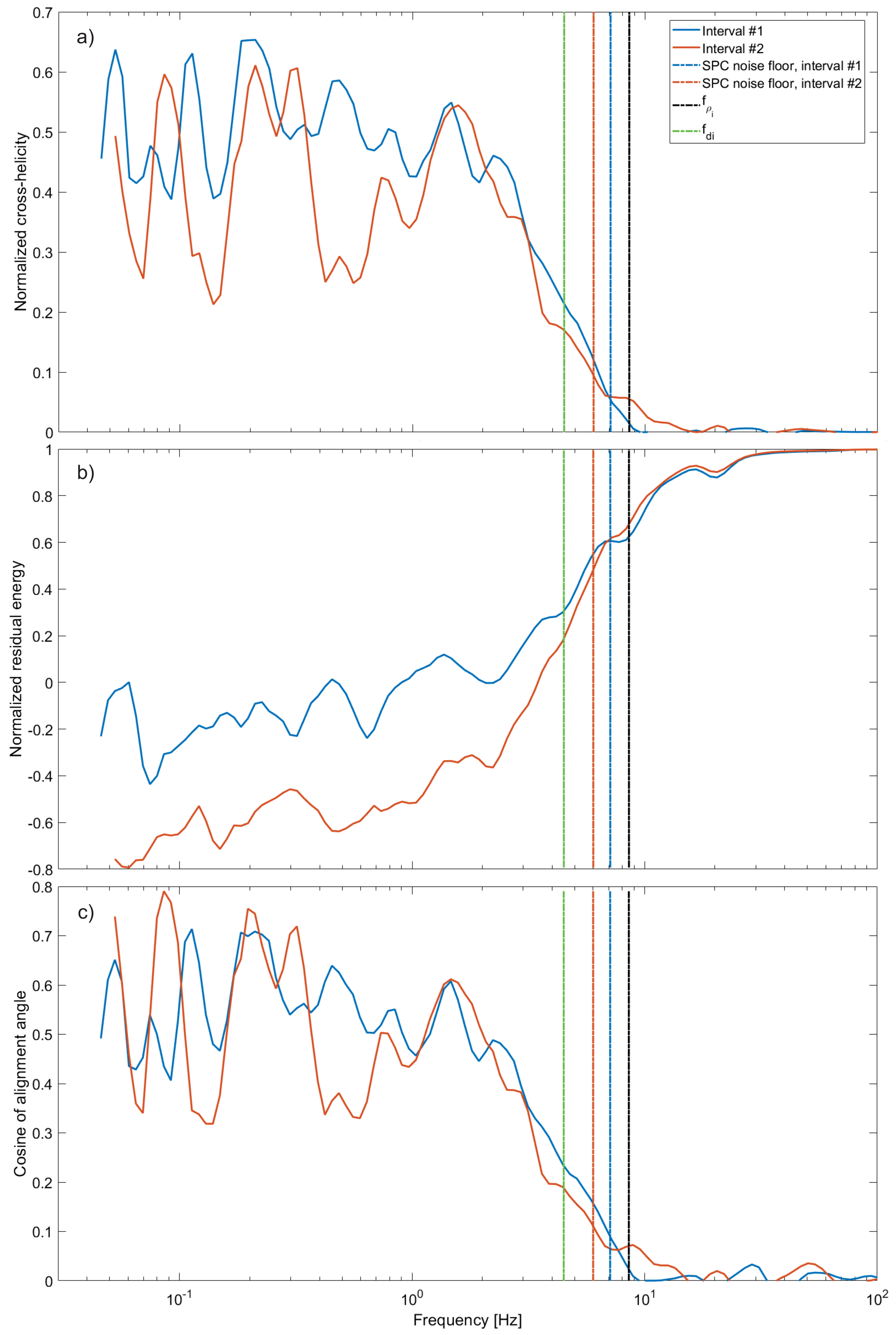}
\caption{Normalized cross helicity, residual energy and cosine of the alignment angle for the first and second FA mode intervals, respectively}
  \label{fig:2}
\end{figure}


\section{Conclusion} \label{sec:conclusion}

In this Paper, we presented the first results from the Flux Angle operation mode of the Faraday Cup instrument onboard Parker Solar Probe. This operation mode allows rapid (up to 293 Hz) measurements of phase space density fluctuations close to the peak of the proton velocity distribution function. We described an approach to convert the measured phase space density fluctuations into vector velocity components, which were found to be reliable up to 7 Hz, which was above the ion-scale spectral break of the magnetic spectrum. 

In the inertial range the velocity and magnetic power spectras were similar to the observations at 1 AU, at kinetic scales the magnetic power spectra steepened (spectral index was -2.5 / -3) while the velocity power spectra showed no clear break, which is rarely observed at 1 AU. The scaling of $\sigma_c$ and $\sigma_r$ in the inertial range was similar to larger statistical studies at 1 AU \citep{podesta2009scale, parashar2018kinetic, verdini20183d}: signatures of alignment between velocity and magnetic fluctuations was found in the inertial range, however near the ion-scale spectral break (at the scale of 3.1 d$_i$) we found loss of alignment between velocity and magnetic fluctuations, which might be due to demagnetization of protons.

We expect that with decreasing perihelion distance the SPC signal-to-noise ratio will improve nearly one order of magnitude and the FA mode will be used several times each day during encounter allowing us to prepare a statistical study and investigate proton velocity fluctuations beyond ion-kinetic scales.\\

\acknowledgments

The SWEAP Investigation and this publication are supported by the PSP mission under NASA contract NNN06AA01C. The FIELDS experiment on the PSP spacecraft was designed and developed under NASA contract NNN06AA01C. D.V was supported by NASA's Future Investigators in NASA Earth and Space Science and Technology Program Grant 80NSSC19K1430. S.D.B. acknowledges the support of the Leverhulme Trust Visiting Professorship program. Contributions from S.T.B. were supported by NASA Headquarters under the NASA Earth and Space Science Fellowship Program Grant 80NSSC18K1201. K.G. Klein was supported by NASA grant NNX16AM23G. J.C. Kasper was supported by NASA Grant NNX14AR78G. C.H.K.C. is supported by STFC Ernest Rutherford Fellowship ST/N003748/2. The data used in this study are available from November 12, 2019 at the NASA Space Physics Data Facility (SPDF)

\bibliography{main}

\end{document}